\documentclass[aps,twocolumn,floatfix]{revtex4}
\usepackage{epsfig}
\usepackage{graphicx}
\usepackage{dcolumn}
\usepackage{natbib}
\usepackage{amssymb}

\def\dj{\hbox{d\kern-0,347em \vrule width0,3em height1,252ex
depth-1,21ex \kern0,051em}}

\begin{document}

\title{Crack Roughness in the 2D Random Threshold Beam Model}

\author{Phani K.V.V. Nukala}
\affiliation{Computer Science and Mathematics Division,
Oak Ridge National Laboratory, Oak Ridge, TN 37831-6164, USA}
\author{Stefano Zapperi}
\affiliation{CNR-INFM, S3, Dipartimento di Fisica, Universit\`a di
Modena e Reggio Emilia, Via G. Campi 213A,  41100 Modena,
Italy}
\affiliation{ISI Foundation, Viale S. Severo 65, 10133 Torino,
Italy}
\author{Mikko J. Alava}
\affiliation{Department of Engineering of Physics, Helsinki
University of Technology, FIN-02015 HUT, Finland}
\author{Sr{\dj}an \v{S}imunovi\'{c}}
\affiliation{Computer Science and Mathematics Division,
Oak Ridge National Laboratory, Oak Ridge, TN~37831-6164, USA}

\begin{abstract}
We study the scaling of two-dimensional crack roughness using large
scale beam lattice systems. Our results indicate that the crack
roughness obtained using beam lattice systems does not exhibit
anomalous scaling in sharp contrast to the simulation results
obtained using scalar fuse lattices. The local and global roughness
exponents ($\zeta_{loc}$ and $\zeta$, respectively) are equal to
each other, and the two-dimensional crack roughness exponent is
estimated to be $\zeta_{loc} = \zeta = 0.64 \pm 0.02$. Removal of
overhangs (jumps) in the crack profiles eliminates even the minute
differences between the local and global roughness exponents.
Furthermore, removing these jumps in the crack profile completely
eliminates the multiscaling observed in other studies. We find that
the probability density distribution $p(\Delta h(\ell))$ of the
height differences $\Delta h(\ell) = [h(x+\ell) - h(x)]$ of the
crack profile obtained after removing the jumps in the profiles
follows a Gaussian distribution even for small window sizes
($\ell$).
\end{abstract}
\maketitle

\section{Introduction}
Understanding the scaling properties of fracture surfaces still represents 
an unsolved problem despite two decades of intense research
activities \cite{breakdown,alava06}.
Experiments on several materials under different loading conditions
have shown that the fracture surface is self-affine \cite{man} and
can be characterized by a roughness exponent $\zeta$. 
Experiments on several materials including
metals \cite{metals}, glass \cite{glass}, rocks \cite{rocks} and
ceramics \cite{cera}, have shown a universal out of plane roughness exponent of
$\zeta \simeq 0.8$ for three-dimensional fracture surfaces irrespective of
the material studied, as revieved in Ref.
\cite{bouch}. Recent experimental evidence shows, however, that the picture is more
complicated: such scaling is valid at small and intermediate scales
in the so-called fracture process zone (FPZ), while at large scales
one observes a new regime with $\zeta\simeq 0.4$ attributed to crack line depinning
\cite{ponson06,bonamy06,ponson06ijf}. Recent numerical investigations of the
random fuse model indicate that the local roughness exponent in two
dimensions does not depend on the size of the FPZ, but only on the
fact that a FPZ is present or not \cite{nukala07}.

In addition, it is by no means a priori clear that "simple"
self-affinity is sufficient to describe the experiments.  
It has been argued that fracture surfaces may
exhibit anomalous scaling \cite{anomalous}: the {\it global}
exponent describing the scaling of the crack width with the sample
size is larger than the local exponent measured on a single sample
\cite{exp-ano,exp-ano2}. In this sense, it is necessary to introduce
two roughness exponents a global exponent ($\zeta$) and a local
exponent ($\zeta_{loc}$) to define the roughness of fracture
surfaces. Anomalous scaling is noted in the numerical simulations as
well \cite{zapperi05,nukalapre3D}; however, its origin is not clear
yet although in experiments it is conjectured to be an artifact of
initial transient regime as the fracture front moves away from the
initial notch \cite{ponson06ijf}. Recent studies
\cite{jstat2,bakke07} also suggest that the origin of anomalous
scaling in numerical simulations in two dimensions may also be due
to the existence of overhangs (jumps) in the crack profile,
originating from crack branching. As a further complication, there is
an on-going debate whether fracture surfaces exhibit multi-affine scaling 
\cite{procaccia,jstat2,bakke07,santucci07}, implying that
one would observe a whole family of roughness
exponents $\zeta_q$ depending on which statistical moment $q$ of
the correlation function is measured.

The theoretical understanding of the origin and universality of
crack surface roughness is often investigated by discrete lattice
(fuse, central-force, and beam) models \cite{alava06}. In these models the elastic
medium is described by a network of discrete elements such as fuses,
springs and beams with random failure thresholds. In the simplest
approximation of a scalar displacement, one recovers the random fuse
model (RFM) where a lattice of fuses with random threshold are
subject to an increasing external voltage \cite{deArcangelis85}.
Using two-dimensional RFM, the estimated crack surface roughness
exponents are: $\zeta = 0.7\pm0.07$ \cite{hansen91b},
$\zeta_{loc}=2/3$ \cite{sep-00}, and $\zeta = 0.74\pm0.02$
\cite{bakke}. Recently, using large system sizes (up to $L = 1024$)
with extensive sample averaging, we found that the crack roughness
exhibits anomalous scaling \cite{zns05}.
In particular, the local and global roughness exponents estimated
using two different lattice topologies are: $\zeta_{loc} =
0.72\pm0.02$ and $\zeta = 0.84\pm0.03$. The reasons behind the
origin of anomalous scaling in numerical simulations are not yet
clear, although the existence of overhangs in the crack profile is
expected to have contributed to anomalous scaling
\cite{jstat2,bakke07}. In comparison, the roughness exponents
obtained from quasi two-dimensional experiments, mainly obtained for
paper samples, indicate a roughness exponent in the range $\zeta
\simeq 0.6-0.7$ \cite{jstat2,santucci07,kertesz93,engoy94,salminen03,rosti01}, 
but occasionally significantly higher values have also been 
reported \cite{menezessobrinho05}. 
It is not known at this time whether this variation in $\zeta$
values is a reflection of practical difficulties in experimentally
measuring $\zeta$ - in paper it is difficult to have a scaling range 
spanning over several decades since the structure becomes
three-dimensional at small scales (0.1 mm) and at the millimeter
range the fiber length interferes - or that the roughness exponent
is not really universal but depends on material parameters and the
anisotropy of the medium.

Despite this reasonable agreement between the numerical results
obtained using two-dimensional RFM and the above
quasi-two-dimensional experimental results, a lingering question is
whether scalar representation of the elastic medium using random
fuse models is an adequate representation of fracture. Moreover, it
is an intriguing question whether the same roughness exponents as
those obtained using RFM will be obtained using more complex random
threshold central-force (spring) and beam models.  Using
central-force models, a roughness exponent of $0.65 \pm 0.07$ is
obtained in Ref. \cite{mala06}. A more recent study with a range of
disorder strengths in random thresholds estimated the local and
global roughness exponents to be in the range $0.5-0.67$ and
$0.6-0.85$, respectively \cite{bakke07}. 
Using two-dimensional beam
simulations, Ref. \cite{hansenbeam} estimated a typical value for 
the global roughness exponent to be $0.86$; however, the roughness 
exponent is argued to be disorder dependent. 
Similar disorder dependent roughness exponents have also been 
reported recently for the RFM \cite{hansennew}. These results 
were obtained in the square lattice and were attributed to a 
lattice effect at low disorder. They indeed disappear for
triangular lattice, where the roughness exponent is the same
independent on disorder \cite{nukala07}.

The questions we address in this article are the following:
(i) whether anomalous scaling is present in two-dimensional fracture simulations
using beam lattice systems, and (ii) whether roughness exponents so measured using beam
lattice systems are in agreement with those obtained using simplified scalar RFM models.
Along the way, we also address why scalar RFM models have been successful in
representing fracture in a disordered elastic medium. Recent studies
\cite{salminen03,jstat2,santucci07} have shown that multi-affinity of fracture surfaces
is an artifact of overhangs (or jumps) in the crack profile. Here, we further
investigate the influence of these overhangs in the crack profiles on crack roughness exponents.
This article has three further Sections: in the next one, we
describe the beam model used. In Section III, we present the
numerical results obtained with it. Finally, Section IV concludes the
paper.

\section{Beam Model}
The random thresholds beam model (RBM) we consider in this study is
a two-dimensional triangle lattice system of size $L \times L$.
Unlike the scalar RFM model, the vectorial RBM has three degrees of
freedom (x-translation $u$, y-translation $v$, and a rotation
$\theta$ about z axis) at each of the lattice nodes (sites), and
each of the bonds (beams) in the lattice connects two nearest
neighbour nodes. We assume that the beams are connected rigidly at
each of the nodes such that the angle between any two beams
connected at a node remains unaltered during the deformation
process. These nodal displacements and rotations introduce conjugate
forces and bending moments in the beam members. Using Timoshenko
beam theory \cite{przemieniecki}, which includes shear deformations
of the beam cross-section in addition to the usual axial deformation
of cross-sections, the {\it local} stiffness matrix for a beam
element that relates the local nodal displacements and rotations to
local nodal forces and bending moments in the beam's {\it local}
coordinate system is given by
\begin{eqnarray}
{\bf K}_{local} & = & \left[\begin{array}{ccccccccc}
\frac{EA}{\ell_b} & 0 & 0 & -\frac{EA}{\ell_b} & 0 & 0 \\
 & \frac{12EI}{(1+\alpha) \ell_b^3} & \frac{6EI}{(1+\alpha) \ell_b^2} & 0 & -\frac{12EI}{(1+\alpha) \ell_b^3} & \frac{6EI}{(1+\alpha) \ell_b^2} \\
 & & \frac{(4+\alpha) EI}{(1+\alpha) \ell_b} & 0 & -\frac{6EI}{(1+\alpha) \ell_b^2} & \frac{(2-\alpha) EI}{(1+\alpha) \ell_b} \\
 & & & \frac{EA}{\ell_b} & 0 & 0 \\
 & SYM & & & \frac{12EI}{(1+\alpha) \ell_b^3} & -\frac{6EI}{(1+\alpha) \ell_b^2} \\
 & & & & & \frac{(4+\alpha) EI}{(1+\alpha) \ell_b} \\
\end{array} \right] \label{Klocal}
\end{eqnarray}
where $E$ is the Young's modulus, $G$ is the shear modulus, $A$ is
the beam cross-sectional area, $I$ is the moment of inertia of beam
cross-section, $\ell_b$ is the length of the beam, and $\alpha =
\frac{12EI}{GA \ell_b^2}$ is the shear correction factor, which
denotes the ratio of bending stiffness to the shear stiffness. If
shear deformation of beam cross-section is negligible, then $\alpha
= 0$ and the Timoshenko's beam theory reduces to Euler-Bernoulli
beam theory. Equation \ref{Klocal} presents a relation between local
nodal displacements and rotations ${\bf d}_{\ell} = (u_{li}, v_{li},
\theta_{li}, u_{lj}, v_{lj}, \theta_{lj})^T$ and local forces and
moments ${\bf F}_{\ell} = (F_{li}, V_{li}, M_{li}, F_{lj}, V_{lj},
M_{lj})^T$. In this setting, the subscript $l$ refers to local
quantities, the superscript $T$ represents transpose of a vector or
a matrix, $i$ and $j$ refer to $i$-th and $j$-th nodes of the beam,
and $F$, $V$, and $M$ refer to axial force, shear force, and bending
moments respectively.

Computing the equilibrium of the lattice system is achieved by first
transforming these local quantities (${\bf d}_{\ell}$ and ${\bf
F}_{\ell}$) into global quantities ${\bf d} = (u_{i}, v_{i},
\theta_{i}, u_{j}, v_{j}, \theta_{j})^T$ and ${\bf F} = (F_{i},
V_{i}, M_{i}, F_{j}, V_{j}, M_{j})^T$ through a coordinate
transformation ${\bf T}$ such that ${\bf d}_{\ell} = {\bf T} {\bf
d}$, ${\bf F}_{\ell} = {\bf T} {\bf F}$, and ${\bf K} = {\bf T}^T
{\bf K}_{local} {\bf T}$, and then satisfying equilibrium equations
at each node such that
\begin{eqnarray}
\Sigma_{<ij>} F_x & = & 0; ~~~ \Sigma_{<ij>} F_y = 0; ~~~ \Sigma_{<ij>} M = 0 \label{eqbm}
\end{eqnarray}
where $\Sigma_{<ij>}$ implies that the summation is carried over all the
intact bonds $<ij>$ joining at node $i$. In the above discussion, the transformation
matrix ${\bf T}$ is given by
\begin{eqnarray}
{\bf T} & = & \left[\begin{array}{cccccc}
{\bf Q} & {\bf 0} \\
{\bf 0} & {\bf Q} \\
\end{array} \right] \label{Tarray}
\end{eqnarray}
where
\begin{eqnarray}
{\bf Q} & = & \left[\begin{array}{ccccc}
c & s & 0 \\
-s & c & 0 \\
0 & 0 & 1 \\
\end{array} \right] \label{Qarray}
\end{eqnarray}
and $c = \cos(\beta)$, $s = \sin(\beta)$ refer to the direction cosines of the beam with $\beta$
representing the angle between the beam axis and the $x$-direction.

In this RBM, we start with a fully intact lattice system with beams having unit length, unit square cross-section and
Young's modulus $E = 1$. This results in a unit axial stiffness ($EA/\ell_b = 1$) and
bending stiffness ($12 EI/\ell_b^3 = 1$) for each of the beams in the lattice system.
Since the beam can deform in two independent deformation modes (axial and bending), we
assume randomly distributed bond breaking axial
and bending thresholds, $t_a$ and $t_b$, based on thresholds
probability distributions, $p_a(t_a)$ and $p_b(t_b)$ respectively. The failure criterion for
a beam is defined through an axial force and bending moment
interaction equation (similar to von-Mises criterion in metal plasticity) given by
\begin{eqnarray}
r & \equiv & \left(\frac{F}{t_a}\right)^2 + \frac{\mbox{max}(|M_i|, |M_j|)}{t_b} = 1 \label{inter}
\end{eqnarray}
The beam breaks irreversibly, whenever the failure criterion $r \ge 1$. Periodic boundary
conditions are imposed in the horizontal direction and a constant unit displacement
difference is applied between the top and the bottom of lattice system.

Numerically, a unit displacement, $\Delta = 1$, is applied at the top
of the lattice system and the equilibrium equations (Eq. \ref{eqbm}) are solved to
determine the force in each of the springs. Solution of Eq. \ref{eqbm}
results in global displacements and rotations ${\bf d}$, using which
the local displacements ${\bf d}_{\ell} = {\bf T} {\bf d}$ and the
local forces ${\bf F}_{\ell} = {\bf K}_{local} {\bf d}_{\ell}$ are computed
for each of the intact beams. Subsequently, for each
bond $k$ with nodes $i$ and $j$, the quantities $a_{k} = \left(\frac{F}{t_a}\right)^2$ and
$b_k = \frac{\mbox{max}(|M_i|, |M_j|)}{t_b}$ are evaluated,
and the bond $k_c$ having the smallest value,
\begin{eqnarray}
r_k & = & \frac{-b_k + \sqrt{b_k^2 + 4 a_k}}{2 a_k} \label{req}
\end{eqnarray}
is irreversibly removed (When $a_k = 0$, then $r_k = \frac{1}{b_k}$). The forces
are redistributed instantaneously after a bond is broken implying that
the stress relaxation in the lattice system is much faster than the
breaking of a bond. Each time a bond is broken, it is necessary to
re-equilibrate the lattice system in order to determine the subsequent
breaking of a bond.  The process of breaking of a bond, one at a time,
is repeated until the lattice system falls apart. For the RBM, we
consider a uniform probability distribution in $[0,1]$ for both axial
and bending thresholds disorders.

Numerical simulations of fracture using lattice networks have often
been limited to smaller system sizes due to the high computational
cost associated with solving a new large set of linear equations
every time a new lattice bond is broken. In this work, we use the
multiple-rank sparse Cholesky factorization downdating algorithm
developed by the authors for simulating fracture using discrete beam
lattice systems \cite{nukalajpamg1,nukalaijnme}. For two-dimensional
systems, this low-rank Cholesky factor downdating algorithm is
significantly faster than competing preconditioned
conjugate-gradient based iterative solvers. Using this numerical
algorithm, we were able to investigate fracture in larger lattice
systems (e.g., $L = 320$ in 2D) than those investigated in previous
studies. The lattice system sizes considered in this work are $L =
\{32, 64, 128, 256, 320\}$ with large numbers of sample
configurations, $N_{config} = \{2000, 4000, 640, 400, 200\}$ respectively, in order to reduce the statistical
error in the numerical results.

\section{Crack Roughness}

\begin{figure}[hbtp]
\includegraphics[width=8cm]{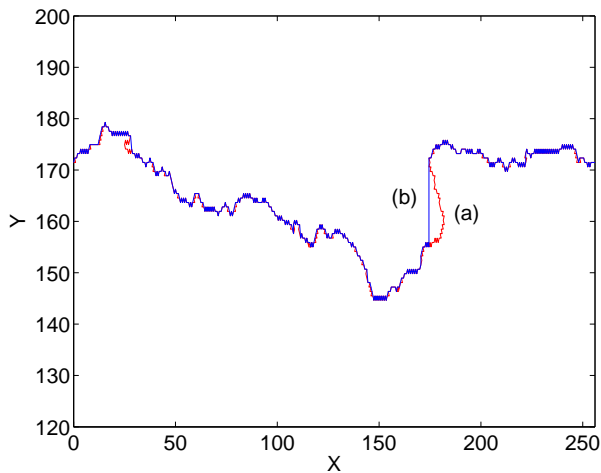}
\caption{(Color online) A typical crack in a system of size $L \times L$
with $L = 256$. This crack, identified as (a) in the figure, has dangling ends, which are removed to obtain a
single valued crack profile $h(x)$, identified as (b) in the figure. However, this final crack $h(x)$
possesses finite jumps that arise due to solid-on-solid projection of fracture surfaces.}
\label{fig:crack}
\end{figure}

Once the sample has failed, we identify the final crack, which typically
displays dangling ends (see Fig. \ref{fig:crack}). We remove them and obtain
a single valued crack line $h_x$, where the values of $x \in [0,L]$.
For self-affine cracks, the local width,
$w(l)\equiv \langle \sum_x (h_x- (1/l)\sum_X h_X)^2 \rangle^{1/2}$,
where the sums are restricted to regions of length $l$ and the average
is over different realizations, scales as $w(l) \sim l^\zeta$
for $l \ll L$ and saturates to a value $W=w(L) \sim L^\zeta$ corresponding
to the global width. The power spectrum
$S(k)\equiv \langle \hat{h}_k \hat{h}_{-k} \rangle/L$, where
$\hat{h}_k \equiv \sum_x h_x \exp i(2\pi xk/L)$, decays as
$S(k) \sim k^{-(2\zeta+1)}$.
When anomalous scaling is present \cite{anomalous,exp-ano,exp-ano2},
the exponent describing the system size dependence of the
surface {\it differs} from the local exponent measured for a fixed system
size $L$. In particular, the local width scales as
$w(\ell) \sim \ell^{\zeta_{loc}}L^{\zeta-\zeta_{loc}}$, so that the global
roughness $W$ scales as $L^\zeta$ with $\zeta>\zeta_{loc}$. Consequently, the
power spectrum scales as $S(k) \sim k^{-(2\zeta_{loc}+1)}L^{2(\zeta-\zeta_{loc})}$.

Recently, Bouchbinder et al. \cite{procaccia} have suggested that
the crack line $h(x)$ is not self-affine; instead, it exhibits a
much complicated multi-affine (or multiscaling) structure. This
implies a non-constant scaling exponent $\zeta_q$ for the $q$-th
order correlation function $C_q(\ell) = \langle
|h(x+\ell)-h(x)|^q\rangle^{1/q} \sim \ell^{\zeta_q}$
\cite{procaccia}. This would imply the breakdown of self-affinity.
Recent studies \cite{salminen03,jstat2,santucci07} have shown that
multi-affinity of fracture surfaces seems to disappear on large
enough scales $\ell$, however.

\begin{figure}[hbtp]
\includegraphics[width=8cm]{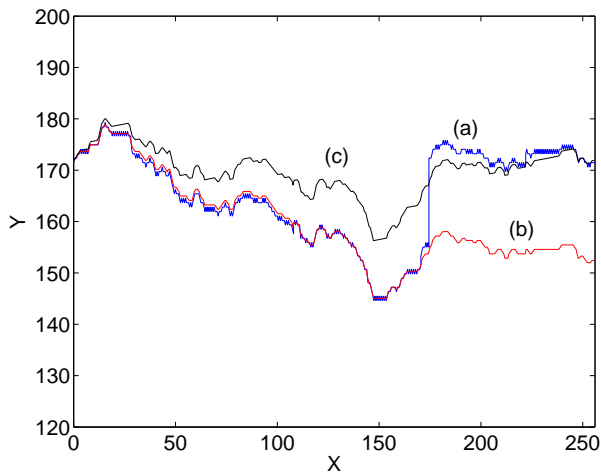}
\caption{(Color online) Figure shows a typical single valued crack profile $h(x)$ with jumps based on
solid-on-solid projection scheme (identified as (a)). Removing the jumps in the
crack profile $h(x)$ makes it a non-periodic profile (identified as (b)). Subtracting a
linear profile from this non-periodic profile results in a periodic profile (identified as (c)).}
\label{fig:crack_nojumps}
\end{figure}

As shown in Fig. \ref{fig:crack_nojumps}, removal of jumps from an
initially periodic crack profile $h(x)$ makes the resulting crack
profile $h_{NP}(x)$ nonperiodic, where the subscript $NP$ refers to
{\it nonperiodicity} of the profiles. A direct evaluation of the
roughness exponent using these nonperiodic profiles can be made.
However, such an evaluation of roughness exhibits finite size
effects for window sizes $\ell > L/2$ due to nonperiodicity.
Alternatively, the roughness of these resulting profiles can be
evaluated by first subtracting a linear profile $h_{lin}(x) =
\left[h_{NP}(0) + \frac{(h_{NP}(L)-h_{NP}(0))}{L} x\right]$ from the
nonperiodic profile $h_{NP}(x)$, and then evaluating the roughness
of the resulting periodic profile $h_{P}(x)$.

\begin{figure}[hbtp]
\includegraphics[width=8cm]{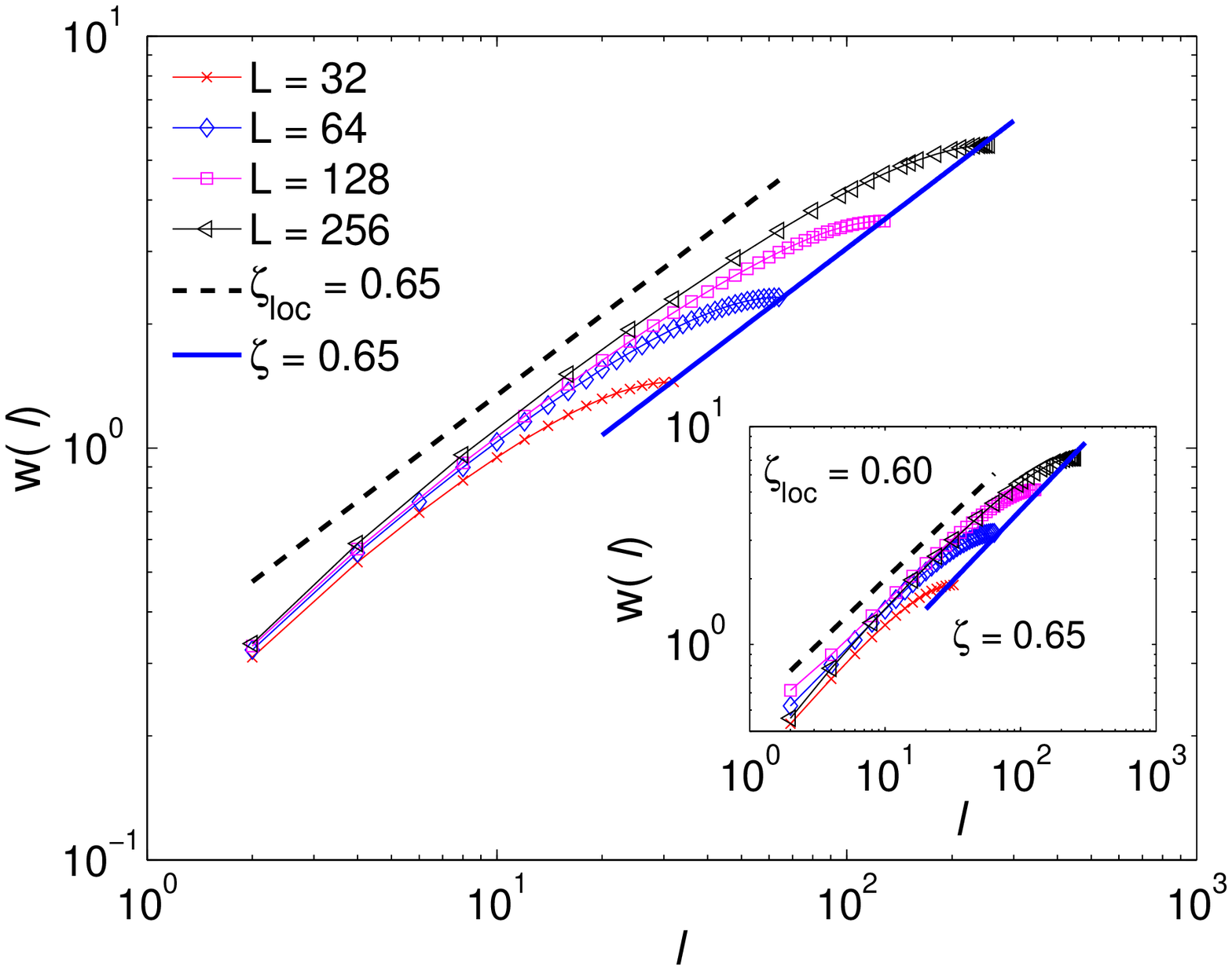}
\includegraphics[width=8cm]{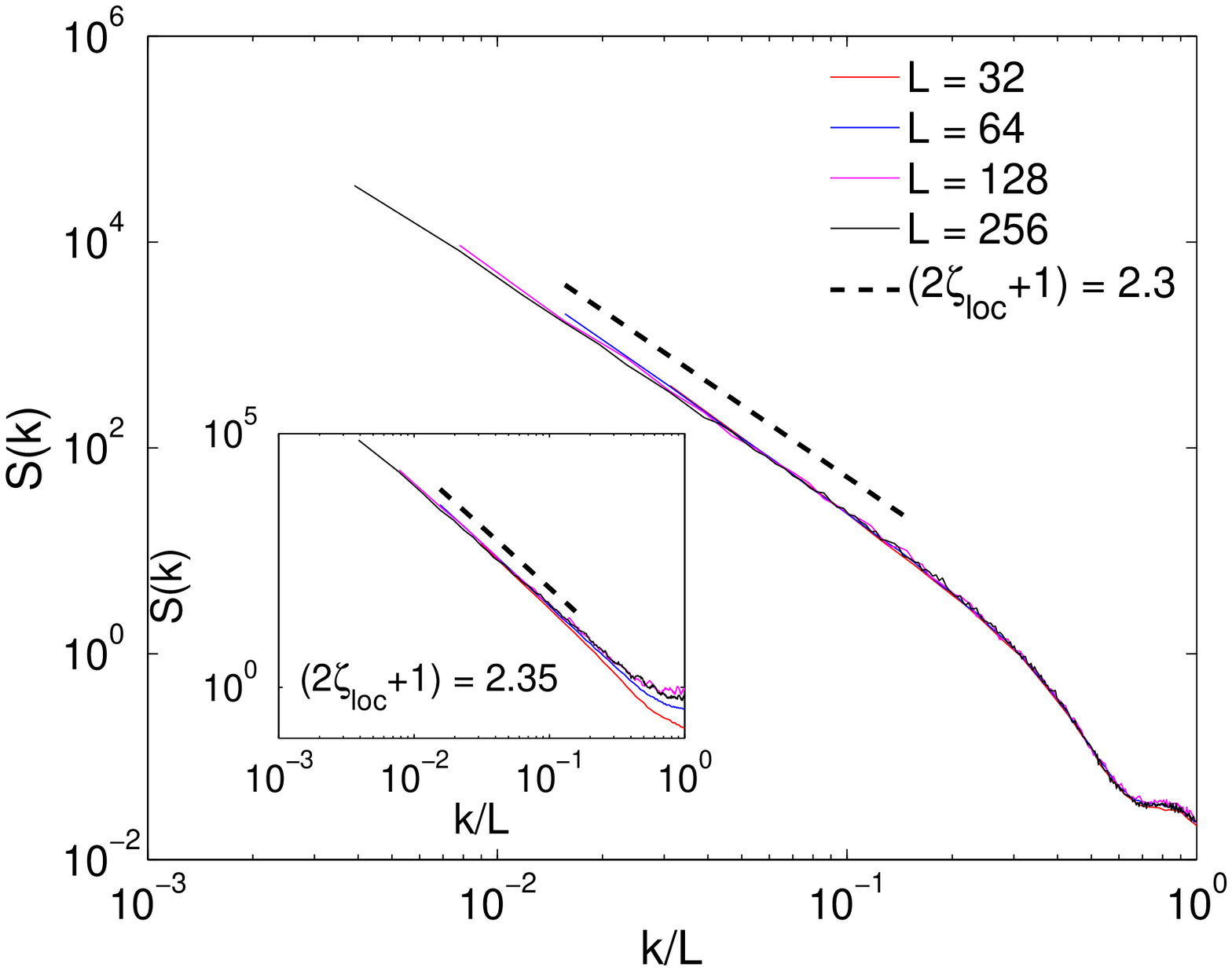}
\caption{(Color online) (a) Scaling of crack width $w(\ell)$ with window size $\ell$. The inset
shows crack widths calculated based on original crack profiles with jumps. The local and
global roughness exponents ($\zeta_{loc}$ and $\zeta$, respectively) appear to be within error bars
although the curves do not collapse.
In the main figure, crack widths are calculated based on crack profiles obtained after
removing the jumps in the profiles. Removing the jumps in the profiles appears to
eliminate the discrepancy between local and global roughness exponents.
(b) Scaling of power spectrum $S(k)$. Inset shows the
scaling of $S(k)$ for original crack profiles whereas the main figure shows the scaling of
$S(k)$ for crack profiles without jumps. Collapse of the power spectrum can be
observed, and the exponents are consistent with those shown in (a).}
\label{fig:movps}
\end{figure}

Figure \ref{fig:movps}a
presents the scaling of crack width $w(\ell)$ with window size $\ell$. The inset and the main figure
respectively show the crack widths calculated based on original crack profiles with jumps and
those obtained from crack profiles without the jumps. The jumps in the
profiles appear to result in slightly different local and
global roughness exponents ($\zeta_{loc}$ and $\zeta$, respectively),
although the exponents are within error bars. However, removing these
jumps in the crack profiles leads to a single roughness exponent of $\zeta_{loc} = \zeta = 0.65$
suggesting that anomalous scaling is an artifact of jumps in the crack profiles, at least for
fracture simulations based on beam lattice systems. We have also investigated the power spectra
$S(k)$ of the crack profiles with and without the jumps in the crack profiles (see Fig. \ref{fig:movps}b).
Collapse of the power spectra for different system sizes can be observed for both the
sets of crack profiles and the roughness exponents ($\zeta_{loc} = \zeta = 0.65$)
obtained using the power law fits to the data are consistent with those
presented in Fig. \ref{fig:movps}a.

The self-affine property of the crack profiles also implies that the
probability density distribution $p(\Delta h(\ell))$ of the height differences
$\Delta h(\ell) = [h(x+\ell) - h(x)]$ of the crack profile
follows the relation
\begin{eqnarray}
p(\Delta h(\ell)) & \sim &  \langle\Delta h^2(\ell)\rangle^{-1/2} ~f\left(\frac{\Delta h(\ell)}{\langle\Delta h^2(\ell)\rangle^{1/2} }\right) \label{pdelt}
\end{eqnarray}
Noting that periodicity in crack profiles is analogous to
return-to-origin excursions arising in stochastic processes, we
propose the following ansatz for the local width $\langle\Delta
h^2(\ell)\rangle^{1/2}$ in height differences $\Delta h(\ell)$
\begin{eqnarray}
\langle\Delta h^2(\ell)\rangle^{1/2} & = &  \langle\Delta h^2(L/2)\rangle^{1/2} ~\phi\left(\frac{\ell}{L/2}\right) \label{dhell}
\end{eqnarray}
with $\langle\Delta h^2(L/2)\rangle^{1/2} = L^\zeta$. For periodic profiles,
the function $\phi\left(\frac{\ell}{L/2}\right)$ is symmetric about $\ell = L/2$ and
is constrained such that $\phi\left(\frac{\ell}{L/2}\right) = 0$ at $\ell = 0$ and $\ell = L$, and
$\phi\left(\frac{\ell}{L/2}\right) = 1$ at $\ell = L/2$. Based on these conditions, a scaling
ansatz of the form
\begin{eqnarray}
\left[\frac{\langle\Delta h^2(\ell)\rangle^{1/2}}{\langle\Delta h^2(L/2)\rangle^{1/2}}\right]^{1/\zeta_{loc}}
+ \frac{(\ell-L/2)^2}{(L/2)^2} & = & 1
\end{eqnarray}
similar to stochastic excursions or bridges can be proposed
for $\langle\Delta h^2(\ell)\rangle^{1/2}$, which implies a
functional form
\begin{eqnarray}
\phi\left(\frac{\ell}{L/2}\right) & = & \left[1 - \left(\frac{(\ell - L/2)}{L/2}\right)^2\right]^{\zeta_{loc}} \label{dhell1}
\end{eqnarray}
for $\phi\left(\frac{\ell}{L/2}\right)$ that is satisfied to a good
approximation by our numerical results. This scaling ansatz results
in anomalous scaling when $\zeta_{loc} \neq \zeta$. Upon further
simplification, Eq. \ref{dhell1} results in
\begin{eqnarray}
\phi\left(\frac{\ell}{L/2}\right) & = & 4^{\zeta_{loc}} \left(\frac{\ell}{L}\right)^{\zeta_{loc}} \left(1 - \frac{\ell}{L}\right)^{\zeta_{loc}}  \label{dhell12}
\end{eqnarray}
which along with $\langle\Delta h^2(L/2)\rangle^{1/2} = L^\zeta$ and Eq. \ref{dhell} shows
how anomalous scaling arises and how local and global roughness exponents
$\zeta_{loc}$ and $\zeta$ can be computed based on numerical results.

\begin{figure}[hbtp]
\includegraphics[width=8cm]{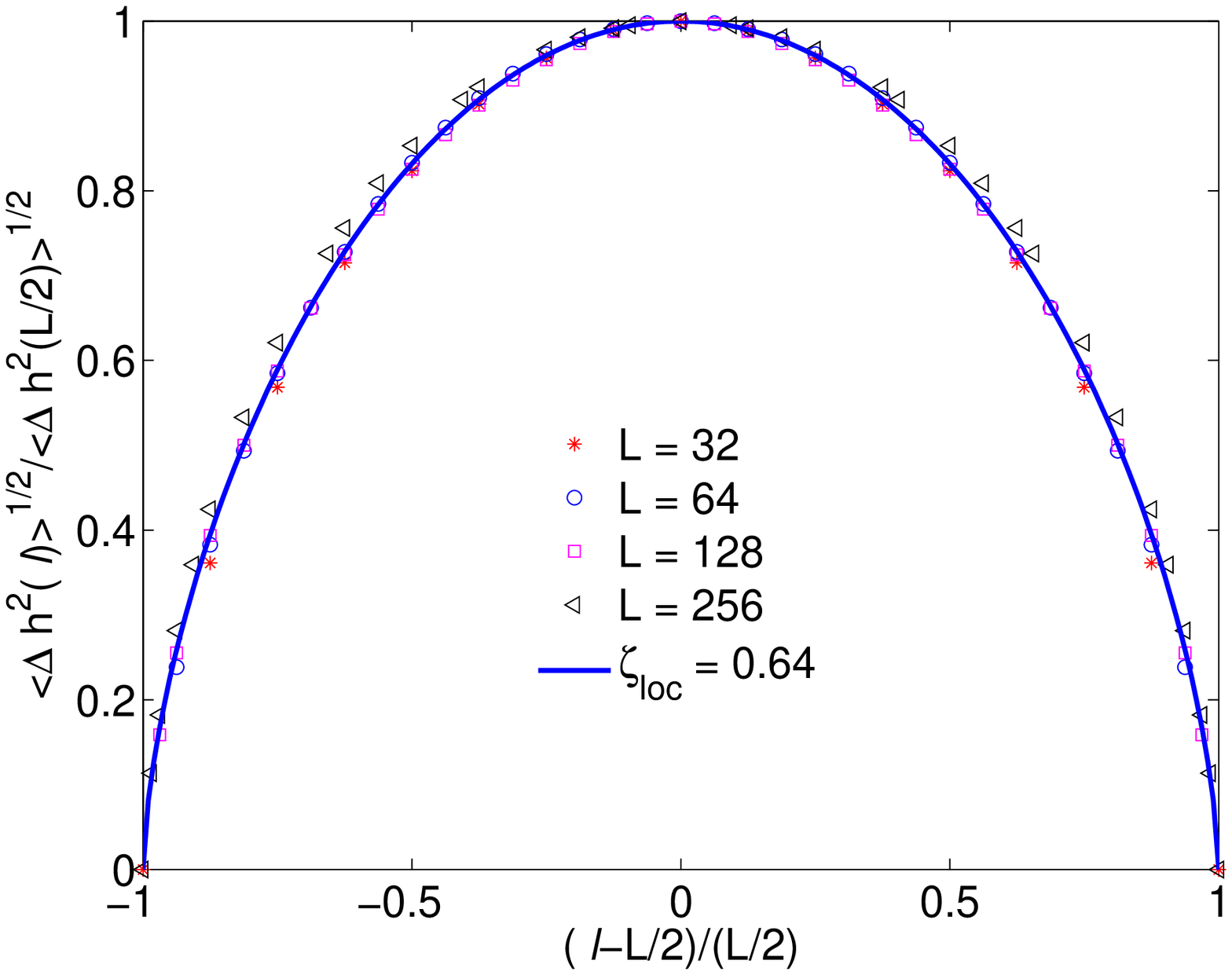}
\includegraphics[width=8cm]{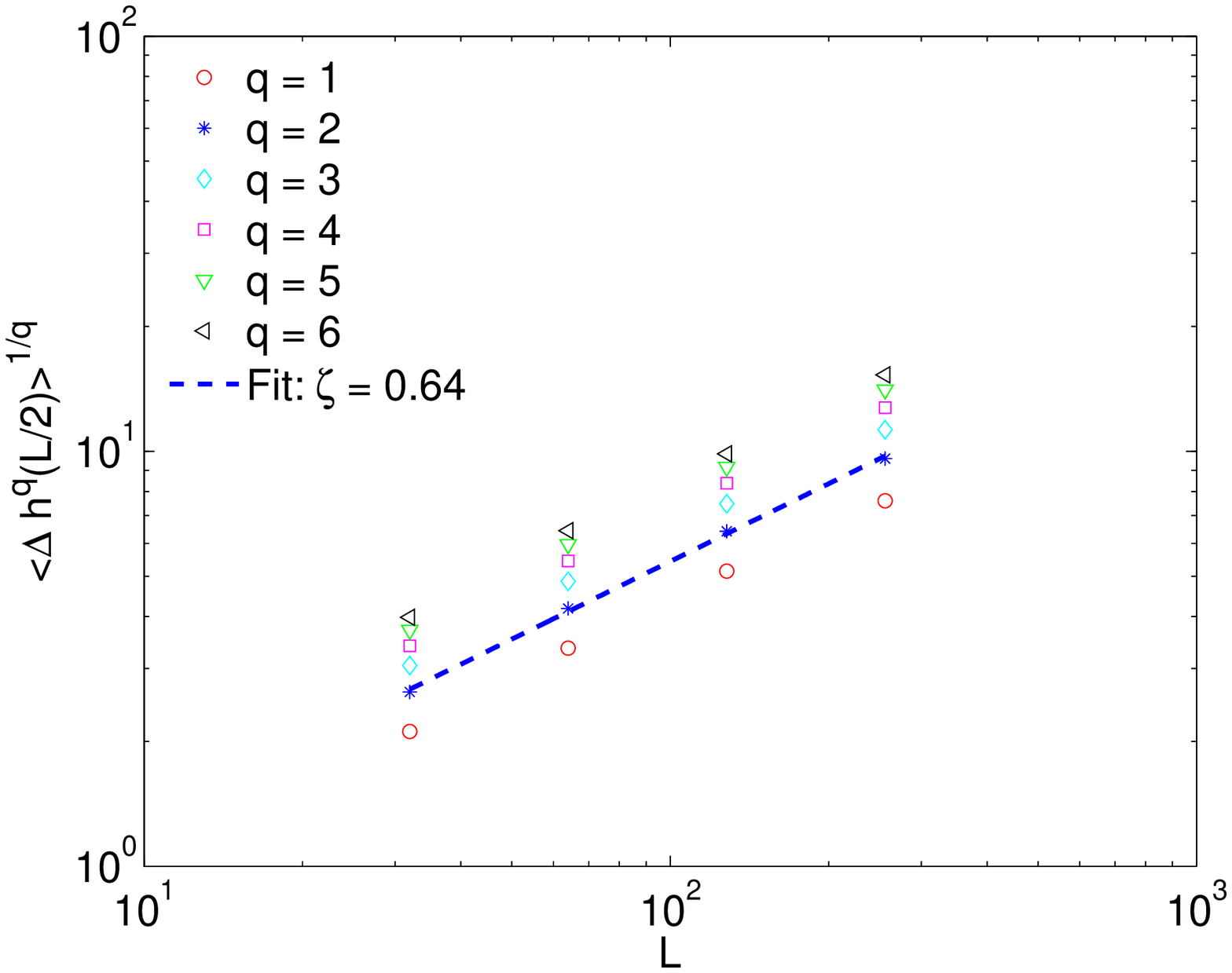}
\includegraphics[width=8cm]{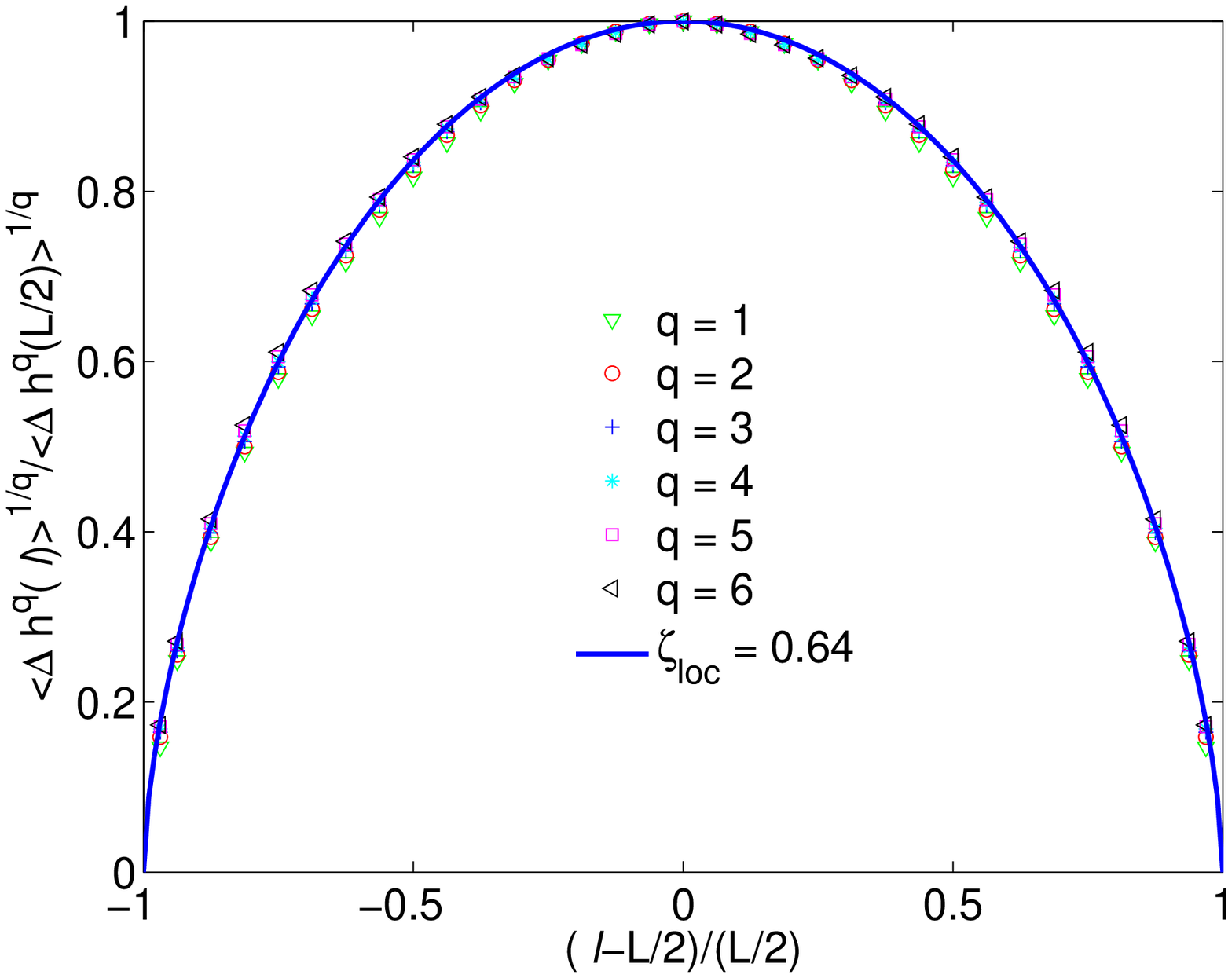}
\caption{(Color online) Removing the jumps from the crack profiles
eliminates anomalous scaling. Furthermore, multiscaling of crack profiles is also due to
the jumps in the crack profiles.
(a) Scaling of $\langle\Delta h_{P}^2(\ell)\rangle^{1/2}$ with window size $\ell$ shows
collapse of data onto Eq. \ref{dhell1} with $\zeta_{loc} = 0.64$. (b) Scaling of
$\langle\Delta h_{P}^2(L/2)\rangle^{1/2}$ with system size $L$ with a scaling
exponent of $\zeta = 0.64$.
(c) Scaling of $\langle\Delta h_{P}^q(\ell)\rangle^{1/q}$ with window size $\ell$
shows that the apparent multiscaling of crack profiles is due to jumps in the profiles.}
\label{fig:pbc_multi}
\end{figure}

Figure \ref{fig:pbc_multi} presents the scaling of $\langle\Delta
h^2(\ell)\rangle^{1/2}$ based on the above ansatz (Eqs. \ref{dhell} and \ref{dhell1}).
The collapse of the $\langle\Delta
h_{P}^2(\ell)\rangle^{1/2}/\langle\Delta h_{P}^2(L/2)\rangle^{1/2}$
data for different system sizes $L$ and window sizes $\ell$ onto a
scaling form given by Eq. \ref{dhell1} with $\zeta_{loc} = 0.64$ can
be clearly seen in Fig. \ref{fig:pbc_multi}(a). In addition, the
collapse of the data presented in Fig. \ref{fig:pbc_multi}(c) for
$\langle\Delta h_{P}^q(\ell)\rangle^{1/q}/\langle\Delta
h_{P}^q(L/2)\rangle^{1/q}$ provides quite concrete evidence that
multi-affine scaling of fracture surfaces, similar to what is
observed in Ref. \cite{procaccia}, is an artifact of jumps in the
crack profile that are formed due to the solid-on-solid
approximation used in extracting the crack profiles. The results in
Fig. \ref{fig:pbc_multi}(c) clearly demonstrate that the removal of
these jumps in the crack profiles completely eliminates this
apparent multiscaling of fracture surfaces. This is also evident
through the scaling of $\langle\Delta h_{P}^q(L/2)\rangle^{1/q}$
presented in Fig. \ref{fig:pbc_multi}(b). The slopes of the data for
moments $q = 1$ to $6$ of $\Delta h_{P}(L/2)$ are identical. It
should also be noted that $\langle\Delta h_{P}^q(L/2)\rangle^{1/q}
\sim L^\zeta$ with $\zeta = 0.64$ and is identical to the local
roughness exponent as obtained from Figs. \ref{fig:pbc_multi}(a) and
(c). This further indicates that anomalous scaling of crack profiles
is not present in fracture simulations obtained using the beam
lattice systems. As already noted in Figs. \ref{fig:movps}a-b, the
difference between the local and global roughness exponents obtained
using the original crack profiles is so small that it already
negates the existence of anomalous scaling of crack roughness using
beam lattice simulations. Removing the jumps caused by overhangs in
the crack profile further reduces even this minute difference in
local and global roughness exponents thereby eliminating anomalous
scaling of crack profiles.

\begin{figure}[hbtp]
\includegraphics[width=8cm]{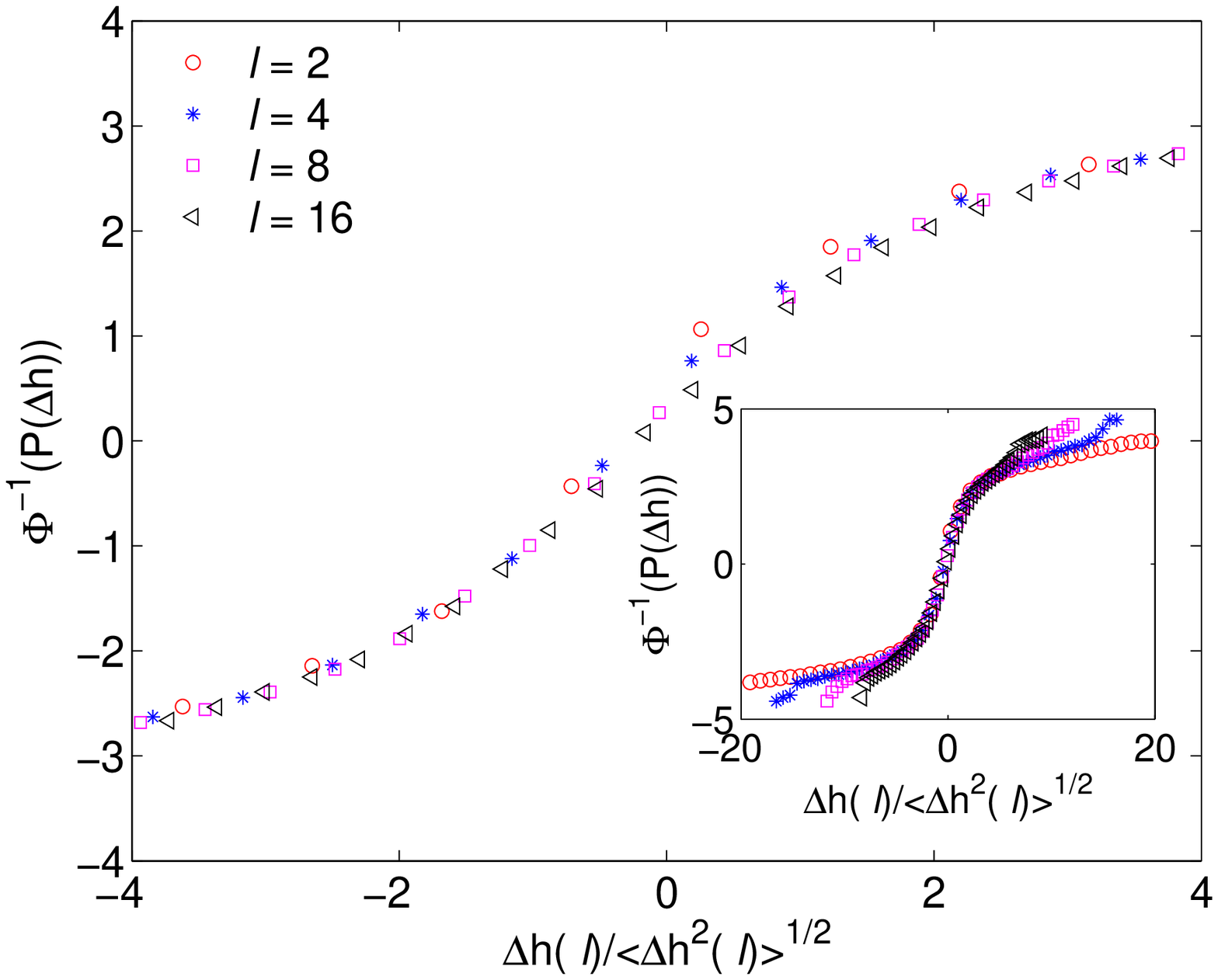}
\includegraphics[width=8cm]{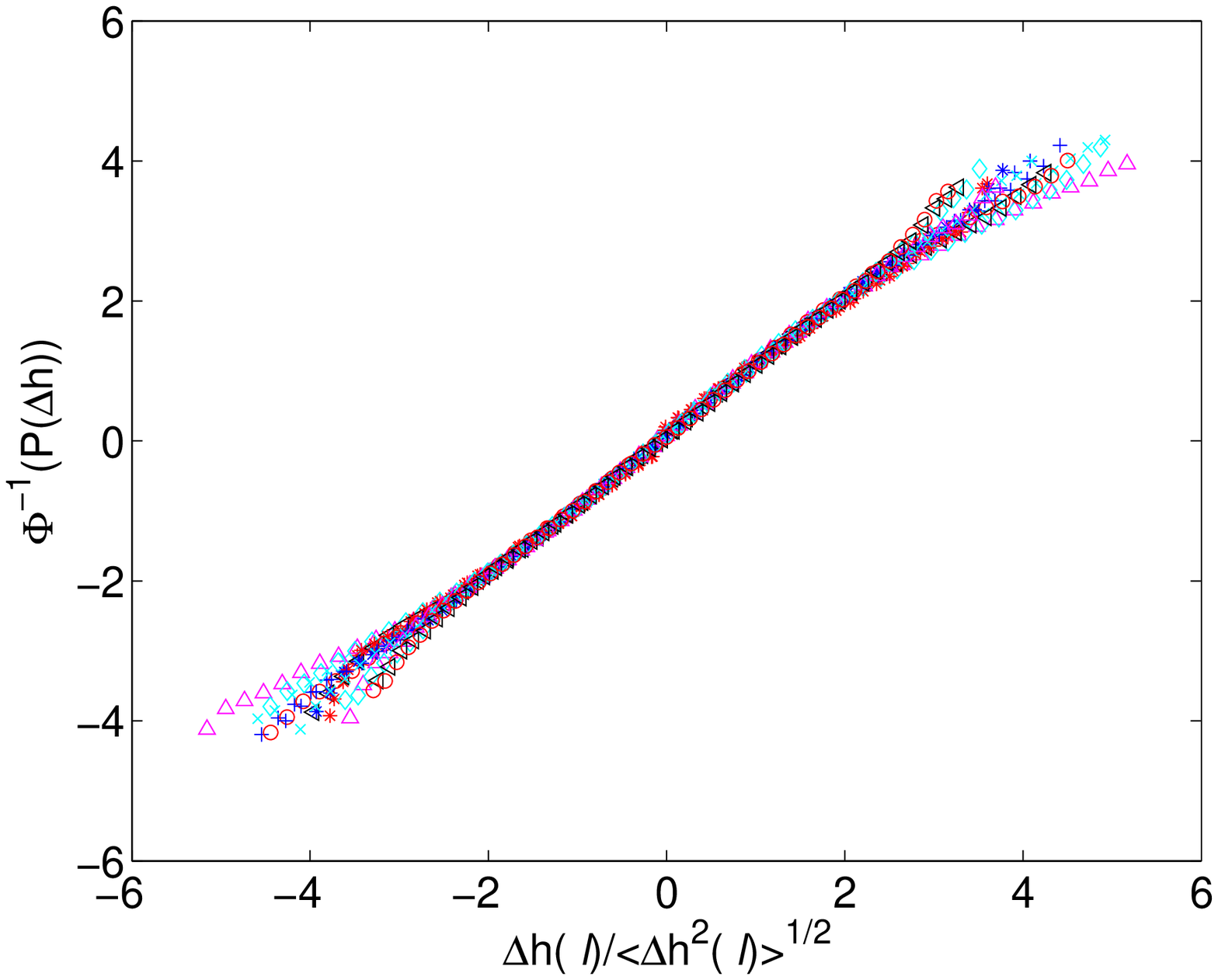}
\caption{(Color online) Normal paper plots of cumulative probability distributions $P(\Delta h(\ell))$
of the height differences $\Delta h(\ell) = [h(x+\ell) - h(x)]$ of the crack profile
$h(x)$ for various bin sizes $\ell$. $\Phi^{-1}$ denotes inverse Gaussian.
The collapse of the profiles onto a straight line with
unit slope would indicate that a Gaussian distribution is adequate to represent $P(\Delta h(\ell))$.
(a) $P(\Delta h(\ell))$ distributions for $L = 256$ and $\ell \ll L$. Large deviation from Gaussian
profiles is observed for these window sizes. (b) Removing the jumps in the
profiles however collapses the $P(\Delta h_{P}(\ell))$ distributions onto a straight line
indicating the adequacy of Gaussian distrbution even for small window sizes $\ell$. The data
is for $L = 128$ with $\ell = (2,4,8,16,32,48,64)$, and
$L = 256$ with $\ell = (2,4,8,16,32,64,96,128)$.}
\label{fig:pdeltah}
\end{figure}

In the following, we investigate the probability density $p(\Delta
h(\ell))$ of height differences $\Delta h(\ell)$. In Refs.
\cite{salminen03,jstat2,santucci07}, the $p(\Delta h(\ell))$
distribution is shown to follow a Gaussian distribution above a
cutoff length scale and the deviations away from Gaussian
distribution in the tails of the distribution have been attributed
to finite jumps in the crack profiles. A self-affine scaling of
$p(\Delta h(\ell))$ as given by Eq. \ref{pdelt} implies that the
cumulative distribution $P(\Delta h(\ell))$ scales as $P(\Delta
h(\ell)) \sim P(\Delta h(\ell)/\langle\Delta
h^2(\ell)\rangle^{1/2})$. Figure \ref{fig:pdeltah}(a) presents the
raw data of cumulative probability distributions $P(\Delta h(\ell))$
of the height differences $\Delta h(\ell)$ on a normal or Gaussian paper for bin
sizes $\ell \ll L$. As observed in Refs.
\cite{salminen03,jstat2,santucci07}, Fig. \ref{fig:pdeltah}(a) shows
large deviations away from Gaussian distribution for these small bin
sizes. However, for moderate $\ell$, the distribution is Gaussian
with deviations in the tails of the distribution beyond the $3\sigma
= 3\langle\Delta h^2(\ell)\rangle^{1/2}$ limit (data not shown in
Figure). Removing the jumps in the crack profiles however collapses
the $P(\Delta h_{P}(\ell))$ distributions onto a straight line (see
Fig. \ref{fig:pdeltah}(b)) indicating the adequacy of Gaussian
distrbution even for small $\ell$. Indeed, Fig. \ref{fig:pdeltah}(b)
shows the collapse of the $P(\Delta h_{P}(\ell))$ data for systems
of sizes $L = 128$ and $L = 256$ with a variety of bin sizes $2 \le
\ell \le L/2$. Removing the jumps in the profiles not only turns the
$P(\Delta h_{P}(\ell))$ distributions Gaussian even for small window
sizes $\ell$ but also extends the validity of $P(\Delta
h_{P}(\ell))$ Gaussian distribution for moderate bin sizes to a
$4\sigma = 4\langle\Delta h_{P}^2(\ell)\rangle^{1/2}$ ($99.993\%$
confidence) limit.

\section{Discussion}

In summary, the analysis and results presented in this paper
indicate that crack profiles obtained in fracture simulations using
the beam lattice systems do not exhibit anomalous scaling of
roughness in contrast with those obtained using the fuse lattice
systems. The local roughness exponent $\zeta_{loc} = 0.65$ obtained
using the beam lattice simulations is in good agreement with the
earlier results obtained using the central-force spring models
($\zeta_{loc} = 0.65$) \cite{mala06} and is rather close to that of
the fuse models $\zeta_{loc} = 0.72$ \cite{zns05}. Notice, however,
that recently it was shown that for the random fuse model on the 
square lattice at low disorder the roughness exponent 
is larger \cite{hansennew}. This result was attributed to bias due to 
the lattice topology. In fact, for the trangular lattice this effect is not
seen and the roughness exponent is universal \cite{nukala07}. Lattice
could also explain the difference between the present result for the
beam model ($\zeta_{loc} = 0.65$) and previous disorder dependent
results obtained for the square lattice \cite{hansenbeam}.

The agreement in the local roughness exponent between the beam, spring, and possibly fuse
models is interesting to note because of these model's dissimilarity
in representing deformation of an elastic medium. This seems to imply
that the anisotropy in the stress redistribution in tensorial models
is irrelevant for the roughness, at least in two dimensions.
We even obtain the same local roughness exponent by
considering a simplified RBM in which failure events form a
connected crack thereby excluding damage nucleation in the bulk.
This simplified beam model is similar to the simplified random
thresholds fuse model (RFM) considered in the Ref. \cite{jstat2}. In
this model, after breaking the weakest beam, successive failure
events are only allowed on beams that are connected to the crack.
Otherwise, the rules of this simplified model strictly follow those
of the usual RBM. Consequently, this model tracks only the connected
crack along with its dangling ends in a disordered medium, and hence
forms the most simplified model to study the effect of disorder on
crack roughness.

\begin{figure}[hbtp]
\includegraphics[width=8cm]{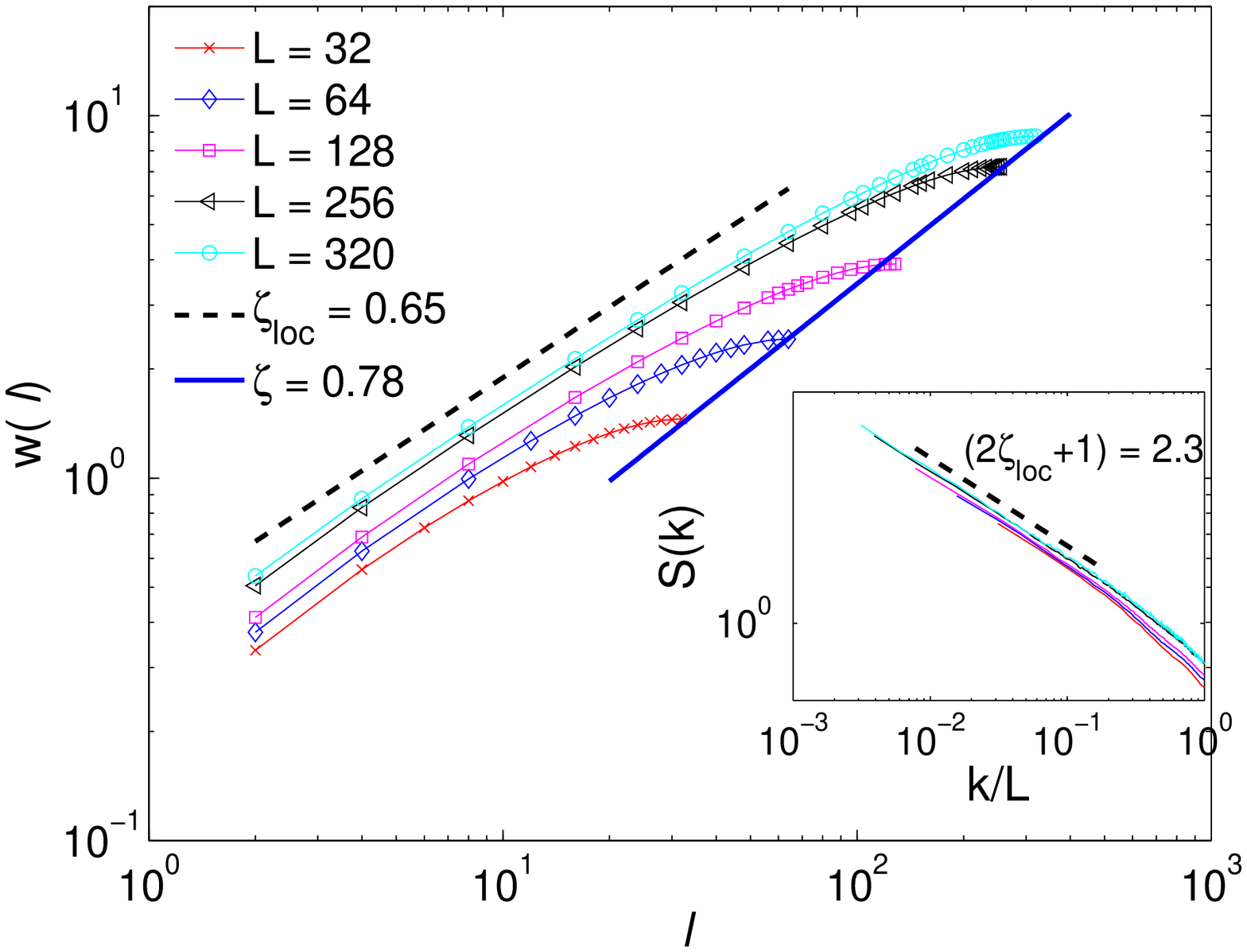}
\includegraphics[width=8cm]{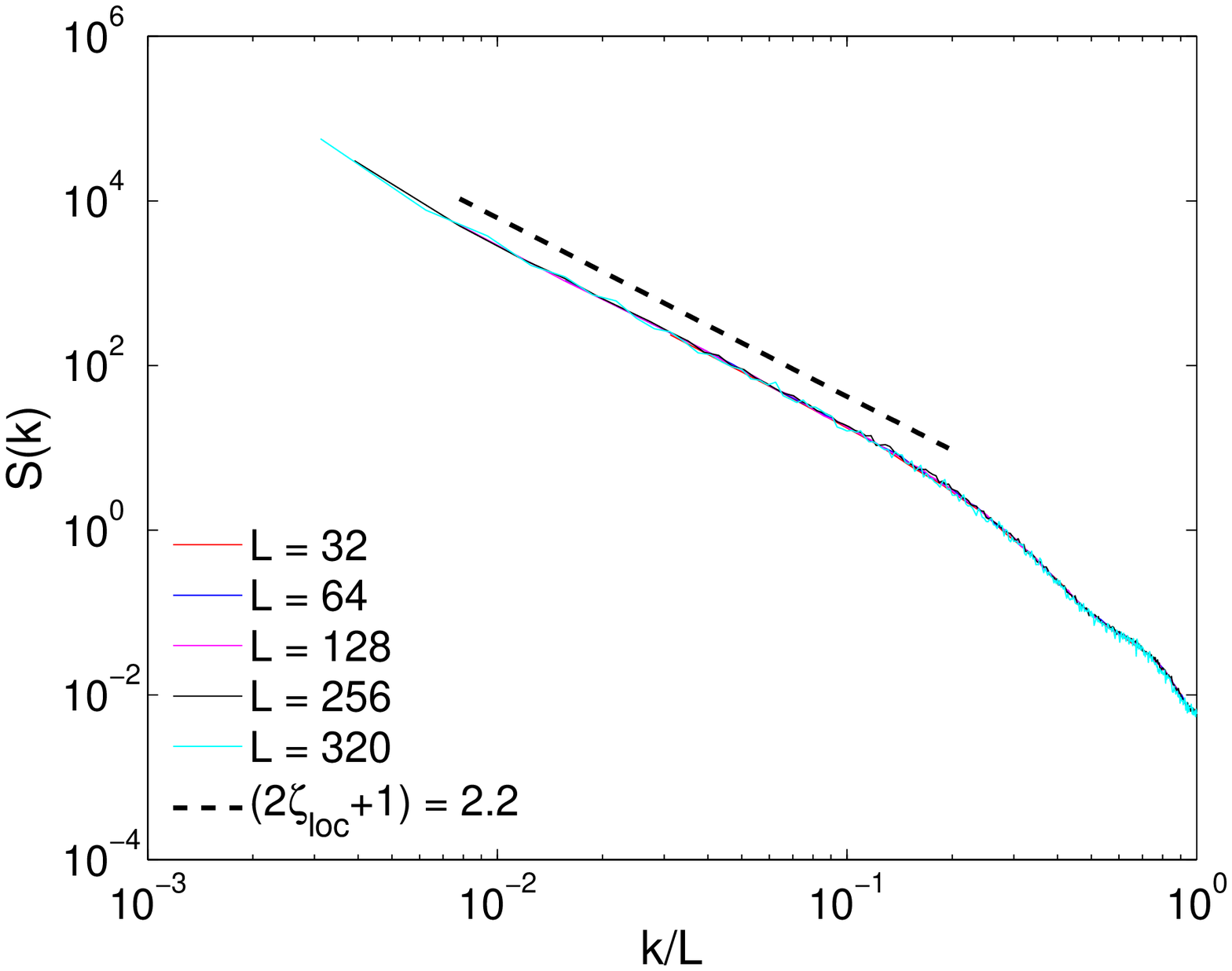}
\caption{(Color online) Scaling of crack widths and power spectra for the simplified RBM model.
(a) Crack profiles (with jumps) seem to exhibit anomalous roughness scaling. However, the
local roughness exponent is same as that noted in conventional RBM. The inset shows the
scaling of power spectra, which are parallel to one another but do not collapse onto each other.
(b) Collapse of power spectra of crack profiles obtained after removing the jumps in the profiles.
Anomalous scaling of roughness in this simplified RBM appears to be due to these jumps in the profiles.}
\label{fig:movps_sing}
\end{figure}

As noted in Ref. \cite{jstat2} for the fuse models, this simplified
beam model exhibits the same local roughness and power spectra
characteristics as that of conventional RBM (see Fig.
\ref{fig:movps_sing}a). This implies that one can expect to obtain
same local roughness exponent as long as there exists a fracture
process zone around the cracks. Additionally, Fig.
\ref{fig:movps_sing}a indicates the existence of anomalous scaling
of roughness as soon as the branching of the cracks is allowed
\cite{remark}. However, removing the jumps in the crack profiles
eliminates this anomalous scaling as can be seen from the collapse
of power spectra shown in Fig. \ref{fig:movps_sing}b. Thus it
appears that the anomalous roughness is two-dimensional fracture
simulations arises due to crack branching and coalescence of
microcracks, which induce jumps in the crack profiles. It should
also be noted that anomalous scaling of crack roughness is readily
evident in  fracture simulations obtained using fuse lattice
systems. The reason for this greater propensity to exhibit anomalous scaling
in fuse lattice systems appears to be due to scalar nature of fuse
systems (anti-planar shear model), which readily allows for crack
branching thereby inducing jumps in crack profiles.

\par
\vskip 1.00em%
\noindent
{\bf Acknowledgment} \\ This research is sponsored by the
Mathematical, Information and Computational Sciences Division, Office
of Advanced Scientific Computing Research, U.S. Department of Energy
under contract number DE-AC05-00OR22725 with UT-Battelle, LLC. MJA and SZ gratefully
thank the financial support of the European Commissions
NEST Pathfinder programme TRIGS under contract NEST-2005-PATH-COM-043386.
MJA also acknowledges the financial support from
The Center of Excellence program of the Academy of Finland.


\begin{thebibliography}{99}
\bibitem{remark}
    As noted in Ref. \cite{jstat2} for the fuse models, when the branching of the cracks or damage within
    the fracture process zone is not allowed thereby limiting the crack extension to only the
    crack tips, a local roughness exponent of $0.5$ is obtained and there is no anomalous scaling.
    As soon as branching is allowed in this simplified RBM model, the
    value of the roughness exponent increases from $\zeta_{loc} = 0.5$ to $\zeta_{loc} \simeq 0.65$.
\end{thebibliography}
\end{document}